\documentclass[aps,showpacs,preprint,prd,superscriptaddress]{revtex4}
\usepackage{latexsym}
\usepackage[T1]{fontenc}
\usepackage[latin9]{inputenc}
\usepackage[letterpaper]{geometry}
\usepackage{float}
\usepackage{amstext}
\usepackage{amsmath}
\usepackage{graphicx}
\usepackage{color}

\makeatletter
\providecommand{\tabularnewline}{\\}

\usepackage{subfig}
\makeatother

\begin{document}

\title{Time Domain Simulations of Arm Locking in LISA}

\author{J.I. Thorpe}
\email{james.i.thorpe@nasa.gov}
\affiliation{Gravitational Astrophysics Laboratory, NASA Goddard Space Flight Center, 8800 Greenbelt Rd., Greenbelt, MD 20771}
\author{P. Maghami}
\affiliation{Attitude Control Systems Engineering Branch, NASA Goddard Space Flight Center, 8800 Greenbelt Rd., Greenbelt, MD 20771}
\author{J. Livas}
\affiliation{Gravitational Astrophysics Laboratory, NASA Goddard Space Flight Center, 8800 Greenbelt Rd., Greenbelt, MD 20771}
\date{\today}

\begin{abstract}
Arm locking is a proposed laser frequency stabilization technique for the
Laser Interferometer Space Antenna (LISA), a gravitational-wave observatory
sensitive in the milliHertz frequency band. Arm locking takes advantage of the
geometric stability of the triangular constellation of three spacecraft that
compose LISA to provide a frequency reference with a stability in the LISA
measurement band that exceeds that available from a standard reference such as
an optical cavity or molecular absorption line. We have implemented a
time-domain simulation of a Kalman-filter-based arm locking system that includes the expected limiting noise sources as well as the effects of imperfect a priori knowledge of constellation geometry on which the design is a based. We use the simulation to study aspects of the system performance that are difficult to capture in a steady-state frequency domain analysis such as frequency pulling of the master laser due to errors in estimates of heterodyne frequency. We find that our
implementation meets requirements on both the noise and dynamic range of the laser
frequency with acceptable tolerances and that the design is sufficiently insensitive to errors in the estimated constellation geometry that the required performance can be maintained for the longest continuous measurement intervals expected for the LISA mission.
\end{abstract}


\maketitle

\section{Introduction}

The Laser Interferometer Space Antenna\cite{Bender_98,Jennrich_09} is a planned
facility for observing gravitational radiation in the milliHertz frequency
band, a regime rich in astrophysical sources. The LISA measurement
concept\cite{Jennrich_09} calls for laser interferometry to be used
to measure fluctuations in the distance between freely-falling test
masses contained within spacecraft separated by $\sim5\times10^{9}\,\mbox{m}$
with a precision of $\sim10\times10^{-12}\,\mbox{m}$, or  $\sim10\,\mbox{pm}$. The interferometric
measurements are performed as a series of one-way measurements
between pairs of spacecraft (SC) and then combined using a technique known
as Time Delay Interferometry (TDI)\cite{Armstrong_99,Shaddock_03}
to form observables that suppress laser frequency noise while retaining
gravitational wave signals. 

The capability of TDI to reject laser
frequency noise is chiefly limited by imperfect knowledge of the absolute
light travel times between the spacecraft (often referred to as the
{}``arm lengths''), which is expected to have an accuracy of $\sim1\,\mbox{m}$ or $\sim 3\,\mbox{ns}$.
With this level of arm length accuracy, the contribution of laser
frequency noise in the TDI observables will satisfy the allocated equivalent
path length noise of $2.5\,\mbox{pm}/\sqrt{\mbox{Hz}}$ so long as
the input laser frequency noise does not exceed a level of

\begin{equation}\tilde{\nu}_{pre-TDI}(f)=\left(282\,\frac{\mbox{Hz}}{\sqrt{\mbox{Hz}}}\right)\cdot\sqrt{1+\left(\frac{2.8\,\mbox{mHz}}{f}\right)^{4}},\label{eq:preTDI-req}\end{equation}

where the Fourier frequency $f$ ranges over the LISA measurement band, $0.1\,\mbox{mHz}\leq f\leq0.1\,\mbox{Hz}$. The expected free-running
noise level of the LISA lasers in the measurement band is roughly $10\,\mbox{kHz}/\sqrt{\mbox{Hz}}\cdot\left(1\,\mbox{Hz}/f\right)$,
which exceeds the requirement in (\ref{eq:preTDI-req}) by more than
four orders of magnitude at the low end of the LISA band. Consequently,
the lasers must be stabilized using an external frequency reference.
A number of candidate stabilization schemes have been studied and
determined to be viable from a noise performance perspective\cite{Shaddock_09,Thorpe10}.
The current focus is on evaluating 
other aspects of
each candidate scheme such as complexity, robustness to implementation errors,
and operational constraints so that the most effective design can be selected.

Two of the candidate schemes rely on arm locking, which utilizes the
existing LISA science signals to derive a frequency reference from
the geometry of the constellation. In one scheme, arm locking is the sole method employed to stabilize the laser frequency where in the other it is combined with another stabilization method in a hybrid system. We focus on the latter case in this paper.

Since its original introduction\cite{Sheard_03},
the arm locking concept has been refined\cite{Sutton_08,McKenzie_09} leading
to improvements in its expected performance. Arm locking has also
been studied in a number of hardware models\cite{Sheard_05,Marin_05,Wand_09},
which have helped to identify potential implementation issues that
were not readily apparent from frequency-domain studies of arm locking.
Chief among these was the discovery that the inability to predict
the heterodyne frequencies of the LISA science signals due to imperfect
knowledge of the inter-spacecraft Doppler shifts leads to {}``frequency
pulling'' of the arm locked laser. If not properly mitigated,
this frequency pulling can be so severe that the laser exceeds
its dynamic range in a matter of hours.

In this paper we present the results from a series of time-domain simulations of  arm locking. The goal is to combine the attractive features of the frequency domain models (realistic noise sources, orbit models, etc.) with those of the hardware models (sensitivity to transients, non-linearities, etc.). The rest of the paper is organized as follows. In section \ref{sec:Background} we describe the problem of arm locking in LISA, defining the relevant signals. In section \ref{sec:Doppler} we briefly review the frequency-pulling effect. We discuss our particular arm locking design and the technical details of our simulation in section \ref{sec:Simulations} and present results in section \ref{sec:Results}.

\section{\label{sec:Background}Arm Locking Model}

To maintain the focus on the arm locking dynamics, we make a few simplifications in our model of the LISA interferometry. The first is that we only consider the interferometric measurements made between different SC (the {}``long-arm'' signals) and ignore the additional measurements made between the SC and the proof mass (the {}``short-arm'' signals).  While a combination of both signals are needed to reach the $\sim10\,\mbox{pm}$ sensitivity levels required for detecting gravitational waves, the $\sim\mbox{nm}$ sensitivity level of the long-arm signals is more than sufficient for frequency stabilization at the level of (\ref{eq:preTDI-req}). 

A second simplification is that we model the two lasers on board the master SC  as a single laser. In reality one of the lasers will be phase locked to the other using a high-gain phase lock loop. The residual noise in this phase lock loop is expected to be far below the other noise sources considered in this paper. Readers interested in additional detail on the LISA interferometric measurement concept should consult one of the many overview papers\cite{Shaddock_08,Jennrich_09,Thorpe10}.

\subsection{Notation}
Our notation system is an adaptation of that used by McKenzie, et al.\cite{McKenzie_09}. The most notable difference is that we represent signals as fluctuations in \emph{frequency} rather than fluctuations in \emph{phase} and replace $\phi$ with $\nu$ to reflect this. 
The three LISA spacecraft (SC) are labeled $SC_{i},\: i=1,2,3$ and
it is assumed that $SC_{1}$ is the master SC. The many different frequency signals are labeled with both an alphabetic and a numeric subscript.  The alphabetic subscript refers to the physical nature of the signal while the numeric subscript refers to the SC involved in producing the signal. In a two digit numeric subscript, the first digit indicates the receiving SC while the second refers to the transmitting SC.  For example, $\nu_{S13}$ denotes the shot noise on the photoreceiver on board $SC_1$ that is receiving signals from $SC_3$. Appendix \ref{sec:notationKey} contains a table summarizing the notation used in this paper and, where possible, the corresponding notation in \cite{McKenzie_09}.

\subsection{A single LISA laser link}

\begin{figure}[H]
\begin{centering}
\includegraphics[width=12cm]{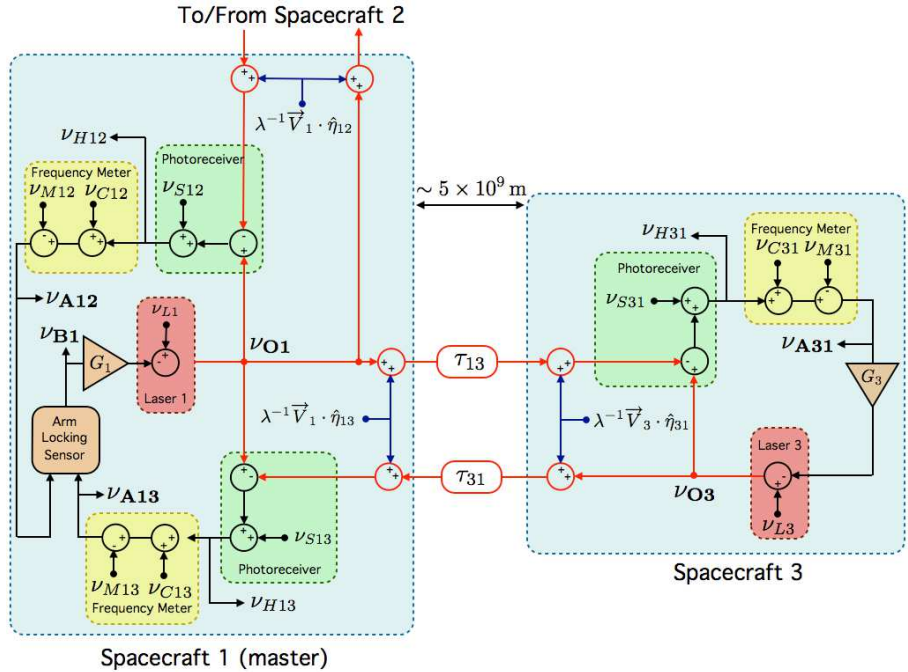}
\par\end{centering}
\caption{\label{fig:AL_diagram}Schematic of frequency signals relevant to
arm locking. See section \ref{sec:Background} of the text for details and Appendix \ref{sec:notationKey} for a key to notation. Adapted from Figure 1 of \cite{McKenzie_09}. }
\end{figure}

Figure \ref{fig:AL_diagram} shows a schematic of the LISA constellation.  We begin our analysis of the arm locking signal chain with Laser 1 on $SC_{1}$, which produces light with a frequency
$\nu_{O1}$. This frequency is a combination of the intrinsic frequency noise
of the laser, $\nu_{L1}$, and the control signal provided by the
arm locking loop. As the laser departs $SC_{1}$ in the direction
of $SC_{3}$, it picks up a Doppler shift due to the motion of $SC_{1}$.
The magnitude of this Doppler shift is $\lambda^{-1}\overrightarrow{V}_{1}\cdot\hat{\eta}_{13}$,
where $\lambda$ is the wavelength of the laser, $\overrightarrow{V}_{1}$
is the velocity of $SC_{1}$, and $\hat{\eta}_{13}$ is the unit vector along the path from $SC_{1}$
to $SC_{3}$. For the purposes of calculating Doppler shifts, we make the assumption that $\hat{\eta}_{ij} = - \hat{\eta}_{ji}$ even though the rotation of the constellation causes these angles to differ on the $\sim\mu$radian level \cite{Jennrich_09}.

The laser then experiences a delay of $\tau_{13}$ on
the order of $5\times10^{9}\,\mbox{m}/c\approx17\,\mbox{s}$ as it
travels to $SC_{3}$. At $SC_{3}$, the signal picks up another Doppler
term due to the motion of $SC_{3}$. At the photoreceiver on $SC_3$ it is interfered with the local laser (with frequency $\nu_{O3}$) to generate an electrical heterodyne signal with frequency 

\begin{equation}\nu_{H31}(t)=\nu_{O3}(t)-\nu_{O1}(t-\tau_{13})-\lambda^{-1}\left[\overrightarrow{V}_{1}(t-\tau_{13})-\overrightarrow{V}_{3}(t)\right]\cdot\hat{\eta}_{13}+\nu_{S31}(t),\label{eq:nuH31}\end{equation}

where $\nu_{S31}$ is a shot noise contribution due to the low light level of the received beam. A device we will refer to as the {}``frequency meter'' (although it is more commonly called a phase meter \cite{Shaddock_06}) is used to measure the frequency of the heterodyne signal. The first step is digitization, which introduces a noise term due to fluctuations in the frequency of the oscillator used to drive the digitizers. To first order, this clock noise is additive with a spectral density that is proportional to the instantaneous heterodyne frequency,

\begin{equation}\tilde{\nu}_{C31}(f)\equiv\nu_{H31}\cdot \tilde{y}_{3}(f).\label{eq:nuC31_def}\end{equation}

Here $\tilde{y}_{3}(f)$ represents the spectrum of fractional frequency fluctuations of the clock on board $SC_3$. 
The frequency meter also measures the frequency of the heterodyne signal relative to some model signal $\nu_{M31}$. The model signal can be used to remove the slow drift of the heterodyne frequency caused by time-varying
Doppler shifts or to impose a constant frequency offset in a phase-lock
loop. The output of the frequency meter is given by

\begin{equation}\nu_{A31}(t)=\nu_{H31}(t) - \nu_{M31}(t) + \nu_{C31}(t).\label{eq:nuA31}\end{equation}

\subsection{Doppler Shifts}

The LISA SC will experience relative velocities along their lines of sight of several meters per second, resulting in Doppler shifts of several MHz.  These Doppler shifts are approximately constant in the LISA measurement band and it is convenient to remove them prior to implementing arm locking. We begin by separating the SC velocity terms into a deterministic term arising from the
SC orbits ($\overrightarrow{V}_{Oi}$) and a stochastic term arising
from attitude jitter of the SC ($\delta\overrightarrow{V}_{i}$)

\begin{equation}\overrightarrow{V}_{i}(t)=\overrightarrow{V}_{Oi}(t)+\delta\overrightarrow{V}_{i}(t).\label{eq:Vel_split}\end{equation}

This in turn leads to two Doppler contributions in the heterodyne signals $\nu_{Hij}$, an orbital motion term ($\nu_{Dij}$) and a SC jitter term ($\nu_{Jij}$) given by 

\begin{equation}\nu_{Dij}(t)=\lambda^{-1}\left[\overrightarrow{V}_{Oi}(t)-\overrightarrow{V}_{Oj}(t-\tau_{ji})\right]\cdot\hat{\eta}_{ij},\label{eq:DopOrbitFreq}\end{equation}

\begin{equation}\nu_{Jij}(t)=\lambda^{-1}\left[\delta\overrightarrow{V}_{i}(t)-\delta\overrightarrow{V}_{j}(t-\tau_{ji})\right]\cdot\hat{\eta}_{ij}.\label{eq:JitterNoise}\end{equation}

With these signal definitions, the heterodyne signal on board $SC_{3}$
can be written 

\begin{equation}\nu_{H31}(t)=\nu_{O3}(t)-\nu_{O1}(t-\tau_{13}+\nu_{D31}(t)+\nu_{J31}(t)+\nu_{S31}(t)\label{eq:nuH31_linear}\end{equation}

\subsection{Phase locking on $SC_2$ and $SC_3$}
Arm locking requires that the slave SC ($SC_2$ and $SC_3$ in our example) operate in a transponder mode, returning to the master SC a copy of the light field that was received. This is accomplished by using a phase-lock loop to control the lasers on board the slave SC. Using $SC_3$ in Figure \ref{fig:AL_diagram} as an example, the controller $G_3$ adjusts the frequency of Laser 3 to minimize the output of the frequency meter, $\nu_{A31}$. In the Laplace domain, the output of Laser 3 will be

\begin{equation}
\nu_{O3}(s)=\frac{G_{3}}{1+G_{3}}\left[\nu_{M31}+\nu_{O1}(s)e^{-s\tau_{13}}-\nu_{D31}(s)-\nu_{J31}(s)-\nu_{S31}(s)-\nu_{C31}(s)\right]+\frac{1}{1+G_{3}}\nu_{L3}(s).\label{eq:SC3_PLL}\end{equation}

Under the assumption of a high-bandwidth phase lock loop, $G_{3}\gg1$,
this simplifies to 

\begin{equation}
\nu_{O3}(s)\approx\nu_{M31}+\nu_{O1}(s)e^{-s\tau_{13}}-\nu_{D31}(s)-\nu_{J31}(s)-\nu_{S31}(s)-\nu_{C31}(s).\label{eq:SC3_PLL_ideal}\end{equation}

\subsection{Formation of the arm locking error signal}

The signal from Laser 3 is transmitted back to $SC_{1}$, picking
up a Doppler contribution form $SC_{3}$, a time delay $\tau_{31}$,
and a Doppler contribution from $SC_{1}$. At $SC_{1}$ it is interfered
with Laser 1 on a photoreceiver, generating shot noise $\nu_{S13}$.
Fluctuations in the heterodyne signal are measured by a frequency
meter, which subtracts a model $\nu_{M13}(t)$ and adds a clock noise
$\nu_{C13}$ to produce the main science signal for the $SC_{1}-SC_{3}$
arm, $\nu_{A13}(t)$. Using (\ref{eq:SC3_PLL_ideal}) to replace $\nu_{O3}$, $\nu_{A13}$ can be written as

\begin{eqnarray}
\nu_{A13}(t) & = & \left[\nu_{O1}(t)-\nu_{O1}(t-\tau_{13}-\tau_{31})\right]+\left[\nu_{J13}(t)+\nu_{J31}(t-\tau_{31})\right]\nonumber \\
 & + & \left[\nu_{S13}(t)+\nu_{S31}(t-\tau_{31})\right]+\left[\nu_{C13}(t)+\nu_{C31}(t-\tau_{31})\right]\nonumber \\
 & + & \left[\nu_{D13}(t)+\nu_{D31}(t-\tau_{31})\right]-[\nu_{M13}(t)+\nu_{M31}(t-\tau_{31})].\label{eq:nu_A13_time}\end{eqnarray}

The second to last bracketed term in (\ref{eq:nu_A13_time}) represents
the deterministic part of the heterodyne signal. The model signal
in the frequency meter, $\nu_{M13}(t)$, can be used to remove
this term, leaving behind a (hopefully small) residual error term,
$\nu_{E13}(t)$, 

\begin{equation}\nu_{M13}(t)=\nu_{D13}(t)+\nu_{D31}(t-\tau_{31})-\nu_{M31}(t-\tau_{31})-\nu_{E13}(t).\label{eq:nuM13}\end{equation}

These residual errors lead to frequency pulling of the master laser.
Section \ref{sec:Doppler} presents estimates for the size of
these errors. With the deterministic terms (mostly) removed, $\nu_{A13}$ can be represented in the Laplace domain as

\begin{eqnarray}
\nu_{A13}(s) & = & \nu_{O1}(s)\left[1-e^{-s(\tau_{13}+\tau_{31})}\right]+\left[\nu_{J13}(s)+\nu_{J31}(s)e^{-s\tau_{31}}\right]+\left[\nu_{C13}(s)+\nu_{C31}(s)e^{-s\tau_{31}}\right]\nonumber \\
 & + & \left[\nu_{S13}(s)+\nu_{S31}(s)e^{-s\tau_{31}}\right]+\nu_{E13}(s).\label{eq:nuA13_s}\end{eqnarray}

The $SC_{1}-SC_{2}$ arm produces a signal, $\nu_{A12}(s)$, that
is analogous to $\nu_{A13}(s)$. The arm locking sensor is a linear
combination of these two signals that is used to estimate
the Laser 1 fluctuations, $\nu_{O1}$, so that they can be suppressed in a feedback loop. The output of the arm locking
sensor, labeled $\nu_{B1}$ in Figure \ref{fig:AL_diagram} is given
by

\begin{equation}\nu_{B1}=\textbf{S}\left[\begin{array}{c}\nu_{A12}\\\nu_{A13}\end{array}\right],\label{eq:nuB1def}\end{equation}

where $\textbf{S}$ is the arm locking sensor vector that describes the specific linear combination of the two individual arm signals. For example,
the {}``common-arm'' sensor, uses the sensor vector $\textbf{S}_{+}\equiv[\frac{1}{2},\:\frac{1}{2}]$.  Table I in \cite{McKenzie_09} provides expressions for several arm locking sensors that have been studied in the literature.

\subsection{Noise Levels}
\label{subsec:noiseLevels}

\subsubsection{Intrinsic Laser Frequency noise}

For this work we assume a pre-stabilized, frequency tunable laser
source with a frequency noise spectral density in the LISA band of

\begin{equation}\tilde{\nu}_{L}(f)=\left(800\,\frac{\mbox{Hz}}{\sqrt{\mbox{Hz}}}\right)\cdot\sqrt{1+(\frac{2.8\,\mbox{mHz}}{f})^{4}}\quad 0.1\,\mbox{mHz}\leq f \leq 0.1\,\mbox{Hz}.\label{eq:MZnoise}\end{equation}

This is representative of the noise-floor of the Mach-Zehnder interferometer
stabilization system\cite{Steier_09} that will fly on LISA Pathfinder \cite{Armano_09},
a LISA technology demonstrator mission.

\subsubsection{Shot Noise}

Shot noise is uncorrelated at each detector and has an equivalent
frequency noise spectrum of

\begin{equation}\tilde{\nu}_{S}(f)=\sqrt{\frac{hc}{\lambda P_{rec}}}\left(\frac{\mbox{Hz}}{\sqrt{\mbox{Hz}}}\right)\left(\frac{f}{1\,\mbox{Hz}}\right),\label{eq:ShotNoise}\end{equation}

where $\lambda=1064\,\mbox{nm}$ is the laser wavelength and $P_{rec}\sim100\,\mbox{pW}$
is the received power. For these numbers, (\ref{eq:ShotNoise}) gives
$\tilde{\nu}_{S}=43\,\mu\mbox{Hz}/\sqrt{\mbox{Hz}}\cdot(f/1\,\mbox{Hz})$.

\subsubsection{Clock Noise}

The spectral density of the fractional frequency variations of the
SC clocks are estimated to be (Table IV of \cite{McKenzie_09})

\begin{equation}\tilde{y}(f)=2.4\times10^{-12}/\sqrt{f}.\label{eq:ClockNoiseLevel}\end{equation}

While LISA will employ a clock-transfer
scheme to correct for differential clock noise between the SC\cite{Klipstein_06}, we
assume that that correction takes place in post processing on the
ground and is not applied to the arm locking error signals on board the SC.

\subsubsection{Spacecraft Jitter\label{sub:Spacecraft-motion}}

The spacecraft jitter noise model is based on simulations of the drag-free
control performance of LISA\cite{Maghami03}. The jitter can be divided
into two orthogonal components in the plane of the LISA constellation
that are independent. Each of these has a position jitter in the LISA measurement band of 

\begin{equation}\delta\tilde{x}(f)=2.5\,\mbox{nm}/\sqrt{\mbox{Hz}}\quad 0.1\,\mbox{mHz}\leq f \leq 0.1\,\mbox{Hz}.\label{eq:JitterNoiseLevel}\end{equation}

Note that due to the fact that the interior angle of the constellation
is not $90\,\mbox{deg}$, the spacecraft jitter contributions from
$SC_{1}$ will be partially correlated in the frequency meter signals $\nu_{A12}(t)$
and $\nu_{A13}(t)$. Finally, we note that it would in principle be
possible to remove the spacecraft jitter by including the {}``short-arm''
interferometers in both the phase-lock error signals in the far SC
and the arm locking error signals in the master SC. This would reduce
the jitter from $\sim\mbox{nm}$ to $\sim\mbox{pm}$
in the LISA band. As with the clock noise correction, this would require
additional on-board processing and is not necessary to reach the pre-TDI
noise requirement specified in (\ref{eq:preTDI-req}).

\section{\label{sec:Doppler}Laser frequency pulling and Heterodyne estimation}

In section \ref{sec:Background}, we explained how Doppler shifts
arising from the SC orbits enter the long-arm frequency meter signals
and how the deterministic parts of the signals are removed using models
of the heterodyne frequency. If we take the expression for the main
science signal of the $SC_{1}-SC_{3}$ arm, $\nu_{A13}$, as expressed in (\ref{eq:nuA13_s})
and consider only the terms resulting from the laser frequency, $\nu_{O1}$,
and the errors in the heterodyne estimate, $\nu_{E13}$, the result
is\[
\nu_{A13}(s)=\nu_{O1}(s)\left[1-e^{-s(t-\tau_{13}-\tau_{31})}\right]+\nu_{E13}(s).\]
If we take the low frequency limit, $s\rightarrow0$, we find that
the first term vanishes. In other words, the signal in $\nu_{A13}$
is insensitive to fluctuations in laser frequency at zero frequency.
The second term, however, is unaffected. The situation is obviously
the same for $\nu_{A12}$ and also for any arm locking sensor formed
as a linear combination of $\nu_{A12}$ and $\nu_{A13}$. If the arm locking
controller has any gain at zero frequency, it will cause the laser
to ramp in an attempt to zero out the heterodyne error terms. This
laser frequency pulling can be mitigated by reducing the arm locking
loop gain below the LISA measurement band, although care must be taken
to ensure that sufficient gain is still present within band. The rate
of pulling is proportional to the error in the estimate of the heterodyne
frequency, hence the design requirements of the control filter will
be driven by the accuracy with which the heterodyne frequency can
be estimated. 

As pointed out by \cite{McKenzie_09}, it is convenient to combine the
heterodyne models from the individual arms into a common and differential
component. For the case where $SC_{1}$ is the master SC, the common and differential heterodyne models are \begin{eqnarray}
\nu_{M+}(t) & \equiv & \nu_{M12}(t)+\nu_{M13}(t)\nonumber \\
\nu_{M-}(t) & \equiv & \nu_{M12}(t)-\nu_{M13}(t)\label{eq:CommDiffDop}\end{eqnarray}
It is also expected that arm locking may require periodic re-acquisition,
either because of an external disturbance (e.g. pointing of the
high-gain antenna) or because some component of the arm locking system
is in a non-desirable operating range (e.g. laser near a longitudinal mode transition,
arms close to equal, etc.). Consequently, the heterodyne frequency
only needs to be estimated for periods on the order of weeks.
For such periods, it is appropriate to use a quadratic model, 

\begin{equation}\nu_{Mx}(t)\approx\nu_{0x}+\gamma_{0x}t+\alpha_{0x}t^{2},\:\:x=(+,-)\label{eq:DopTaylor}\end{equation}

\subsection{Expected Doppler}

\begin{figure}[H]
\begin{centering}
\subfloat[\label{fig:DopComm}Common Doppler parameters]{\includegraphics[width=10cm]{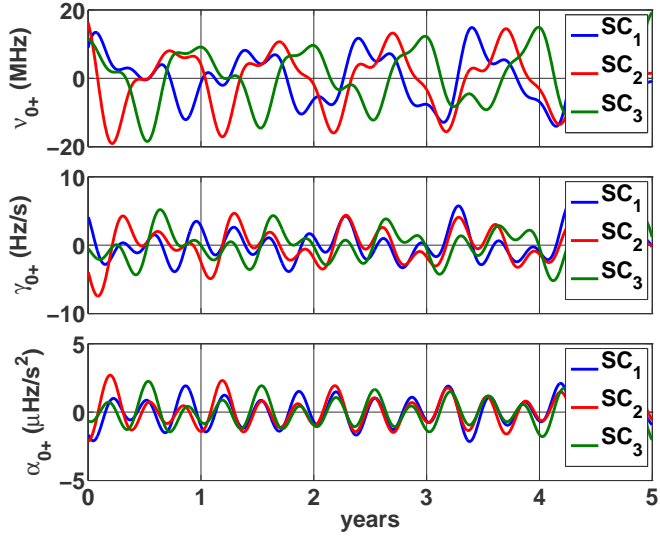}}
\subfloat[\label{fig:DopDiff}Differential Doppler parameters]{\includegraphics[width=10cm]{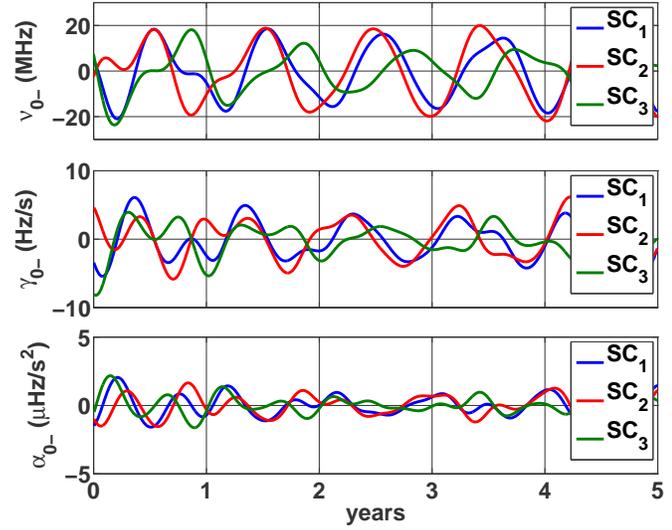}}
\caption{\label{fig:Doppler}Doppler
parameters for the LISA orbital solution \emph{cf3 }in \cite{Hughes_08}.
The plot labels refer to the index of the master spacecraft. }
\end{centering}
\end{figure}

Although the models for the heterodyne signals include both Doppler
shifts and intentional frequency offsets, the Doppler shifts provide
the only source of uncertainty. There are a number of realizations
of the LISA orbits that can be used to derive expected Doppler frequencies.
All exhibit a primary frequency of $\sim 1\,\mbox{yr}^{-1}$ with harmonics
of various amplitudes. There is also a secular component that tends
to degrade the constellation (higher Doppler shifts, larger arm length
mismatches, etc.) as the mission progresses. In Figure \ref{fig:Doppler}
we plot each of the six Doppler parameters in (\ref{eq:DopTaylor}) resulting from an orbital
solution by Hughes \cite{Hughes_08} that was optimized to minimize
the average Doppler frequency in each arm. Each plot contains three
traces, one for each possible choice of master SC.

\subsection{\label{subsec:DoppMeth}Doppler Estimation Methods}

A number of methods have been proposed for determining the Doppler
frequency. One method is to use the orbital ephemeris, such as the
one plotted in Figure \ref{fig:Doppler} to predict the Doppler. With
periodic updates to the ephemeris from ranging data taken during normal
SC down-link operations, the ephemeris velocities can be expected
to be accurate to $3\,\mbox{mm}/\mbox{s}$ \cite{Thornton05}. One issue is that the measured velocity
is the projected component along the line of sight between Earth and
the SC. Transverse velocities are not directly measured but are still
constrained by the orbital model. Consequently one would expect that
the errors in Doppler estimation could differ by a large amount between
different inter-SC links.

Another method for estimating the Doppler frequency is to differentiate
the active ranging signal that is used to determine the absolute link
lengths for the TDI algorithm. Unlike ground tracking,
this method directly measures the velocity along the inter-SC link.
Ranging is expected to have position accuracy of $\sim1\,\mbox{m}$
or better over averaging periods of $\sim1000\,\mbox{s}$ \cite{Esteban_09}. This suggests
that velocities could be measured to $\sim\mbox{mm}/\mbox{s}$ accuracy,
corresponding to Doppler frequency errors on the order of $\sim\mbox{kHz}$.
Additional processing such as longer averaging, Kalman filtering,
or combination with an orbital model may allow for further improvements \cite{Heinzel11}.
In all cases, the processing (including the determination of range
from the pseudo-random code) would take place on ground. Consequently
there would be some delay before the updated Doppler model could be
uploaded to the SC.

McKenzie, et al. \cite{McKenzie_09} proposed a simple method for
determining the heterodyne frequency directly from the frequency meter
data itself. If we consider the expression for the main science signals
(\ref{eq:nu_A13_time}), all of the terms contain mean-zero stochastic
processes with the exception of the Doppler terms. Applying a simple
averaging filter to this signal can suppress the noise terms to reveal
the heterodyne frequency. This simple algorithm relies only on information
from the master SC and could be easily implemented on board. Table \ref{tab:DopErr} gives the errors in the Doppler coefficients estimated by McKenzie, et al. assuming a MZ stabilized laser with a frequency noise spectrum given by (\ref{eq:MZnoise}) and a $200\,\mbox{s}$ averaging time.

\begin{table}[H]
\caption{\label{tab:DopErr}Doppler errors from $200\,\mbox{s}$ averaging of science signal with Mach-Zehnder stabilized laser frequency noise given by (\ref{eq:MZnoise}). Adapted from Table III of \cite{McKenzie_09}.  A * indicates the error was greater than the expected signal and hence the measurement is not used.}
\begin{tabular}{|c|c|c|c|c|c|c|}
\hline 
Parameter & $\nu_{0+}$  & $\nu_{0-}$ &$\gamma_{0+}$ &$\gamma_{0-}$ & $\alpha_{0+}$ & $\alpha_{0-}$ \tabularnewline
\hline
Error & $45\,\mbox{Hz}$ & $0.51\,\mbox{Hz}$  & $2.2\,\mbox{Hz/s}$ & $0.02\,\mbox{Hz/s}$ & * & *\tabularnewline
\hline
\end{tabular}
\end{table}

\section{Arm Locking Simulations\label{sec:Simulations}}

\subsection{\label{sub:Design}Sensor Design}

As mentioned in the introduction, a number of arm locking variants
have been proposed. They differ in the way the science signals from
the two arms extending from the master SC are combined to form an
error signal (and of course the matching controller design that completes the control system). In the language of (\ref{eq:nuB1def}), the sensor vector
$\textbf{S}$ differs for each design. The general goal in designing
the sensor vector is to make the transfer function from laser frequency
noise to arm locking sensor output as simple as possible with flat amplitude and phase responses. Ideally,
$|P(f)|\approx1$ and $\partial \angle P(f)/\partial f = 0$, where 

\begin{equation} P(f)\equiv\frac{\nu_{B1}(f)}{\nu_{O1}(f)}\label{eq:SensorTF}\end{equation}

For example, the original proposal for single arm locking, with sensor
matrix $\textbf{S}_{single}=[\frac{1}{2},\:\:0]$, has $P_{single}(f)=i\sin(2\pi f\tau)e^{-2\pi if\tau}$,
where $\tau=\tau_{12}+\tau_{21}$ is the round-trip light travel time.
This transfer function has nulls at frequencies $f_{n}=n/\tau\approx33\,\mbox{mHz}\cdot n,\: n=1,2,3...$, a number of
which lie in the LISA measurement band. Since the sensor cannot measure
the frequency fluctuations at these frequencies, the control system
cannot correct for them. Furthermore, the rapid swings in the transfer function phase at the $f_{n}$ frequencies make it difficult to design a stable controller that extends beyond $f_1$. More sophisticated arm locking sensors, such
as the modified dual arm locking sensor (MDALS)\cite{McKenzie_09}
make a careful blend of the two science signals to generate a sensor
with a nearly-flat transfer function in the measurement band. 

The problem of blending of multiple sensors to generate the best possible
measurement of a state variable is a classical problem in control
theory. In a previous work\cite{Maghami_09}, we applied Kalman filtering
techniques to generate an arm locking sensor. We will briefly review this approach here.

We begin by making a time-invariant, discrete-time linear state space model of the LISA constellation. The state vector represents the time history of the laser frequency noise over the storage time in the arms,

\begin{equation} \overrightarrow{x}_{k}=\left\{\begin{array}{c}
\nu_{O1}[(k-1)\Delta t] \\
\nu_{O1}[(k-2)\Delta t] \\
\vdots \\
\nu_{O1}[(k-T_{12})\Delta t]
\end{array}\right\}.\label{eq:stateVector}\end{equation}

where $\Delta t$ is the discretization time, $k$ is the time index and $T_{1j}\equiv round [(\tau_{1j} + \tau_{j1})/\Delta t]$ is the index corresponding to the round-trip light travel time between $SC_1$ and $SC_j$.  We assume without loss of generality that $T_{12} > T_{13}$. Note that the first element in the state vector represents the frequency delayed by a single time step as opposed to the instantaneous frequency.

The state vector is updated using the following relations

\begin{equation} \overrightarrow{x}_{k+1} = \mathbf{A}\overrightarrow{x}_k+\mathbf{B}\,u_k + \mathbf{\Gamma}\,w_k,\label{eq:stateAB}\end{equation}

\begin{equation} \mathbf{A}=\left[\begin{array}{cccccc}
0      & 0      & 0      & \cdots    & \cdots  & 0 \\
1      & 0      & 0      & \cdots    & \cdots  & 0 \\
0      & 1      & 0      &\ddots    & \cdots  & \vdots \\
0      & 0      & 1      &\ddots    & \cdots  & \vdots \\
\vdots & \vdots & \ddots &\ddots    & 0       & \vdots \\
0      & 0      & \cdots &\cdots    & 1       & 0 
\end{array}\right]\label{eq:Amatrix}\end{equation}

\begin{equation} \mathbf{B}=\mathbf{\Gamma}=\left[1\:\:0\:\cdots\:0\right]^{T}.\label{Bmatrix}\end{equation} 

The scalar $u_k$ represents the frequency control signal applied to the laser. in this case it would be the output of the block labeled $G_1$ in Figure \ref{fig:AL_diagram}.

\begin{equation} u_k = \nu_{B1}(k\Delta t) \otimes G_1, \label{eq:udef}\end{equation}.

where $\otimes$ denotes convolution. Similarly, the scalar $w_k$ represents the instantaneous frequency noise applied to the laser at time $k \Delta t$,

\begin{equation} w_k = \nu_{L1}(k \Delta t).\label{eq:wdef}\end{equation}

In words, (\ref{eq:stateAB}) says that the laser control signal and the laser noise effect only the first element of the state vector and the remaining elements are determined through simple time delays.

The two frequency meter outputs can be combined to form a two-element measurement vector, $\overrightarrow{y}_k$ which is determined from the following relation,

\begin{equation} \overrightarrow{y}_k\equiv \left[\begin{array}{c} \nu_{A12}(k\Delta t) \\ \nu_{A13}(k\Delta t) \end{array} \right] = \mathbf{C}\,\overrightarrow{x}_k+\mathbf{D}\,u_k+\mathbf{H}\,w_k+\overrightarrow{n}_k.\label{eq:stateMeas} \end{equation}

\begin{equation} \mathbf{C}=\left[\begin{array}{cccccc}
0    & 0  & \cdots & \cdots & \cdots & -1 \\
0    & \cdots & 0 & -1 & 0  & \cdots  \end{array}\right].\label{eq:Cdef}\end{equation}

\begin{equation} \mathbf{D}=\mathbf{H}=\left[ \begin{array}{c} 1 \\ 1 \end{array} \right].\label{Dmatrix}\end{equation} 

$\mathbf{C}$ is a $2\times T_{12}$ matrix describing how the state vector couples into the frequency meter measurements. All elements of $\mathbf{C}$ are zero with the exception of $(1,\,T_{12})$ and $(2,\,T_{13})$, which are $-1$. This represents the fact that the frequency meter measurement includes a copy of the master laser phase delayed by the round-trip light travel time in the arm. The $2\times 1$ vector $\overrightarrow{n}_k$ represents the noise in each of the two frequency meter signals at the current time step. This includes the shot noise, clock noise, and spacecraft jitter noise contributions. The noise level can be determined from the noise spectra in section \ref{subsec:noiseLevels} and the transfer functions to frequency meter outputs in (\ref{eq:nuA13_s}).

Equations (\ref{eq:stateAB}) and (\ref{eq:stateMeas}) are the classical state-space description of a linear system. Kalman filtering \cite{Stengel, Phillips_and_Nagle} is a prescription for generating an optimal estimate of the state vector provided information about the system matrices and noise processes are available. In our case, the state matrices are determined by the arm-lengths (through the definitions of $T_{12}$ and $T_{13}$) and the noise models that determine  $w_k$ and $\overrightarrow{n}_k$. The output of the Kalman filter is an estimate of the state vector, $\overrightarrow{x}_k$, the first element of which corresponds to the laser frequency at time $(k-1)\Delta t$. This element is the output of our Kalman-filter based sensor, which we refer to as an Optimal Arm Locking Sensor or OALS.

\begin{figure}[H]
\begin{centering}
\includegraphics[width=12cm]{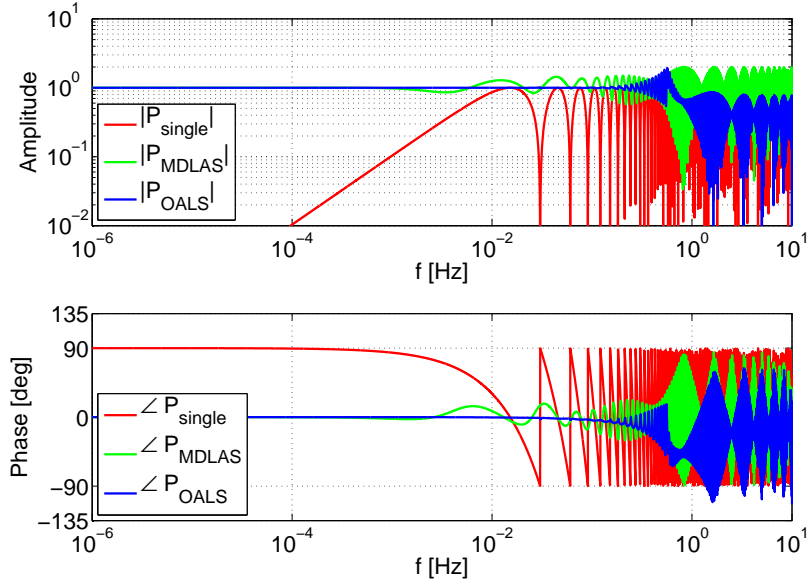}
\par\end{centering}
\caption{\label{fig:SensorComp}Transfer function from master laser frequency noise to sensor output for various arm locking sensors: Single arm sensor\cite{Sheard_03},
MDALS\cite{McKenzie_09}, and OALS (this work). For all cases, the round-trip light travel times in the two arms are $33\,\mbox{s}$ and $32.4\,\mbox{s}$.}
\end{figure}

In Figure \ref{fig:SensorComp}, we show a comparison of the transfer
function from laser frequency noise to sensor output for the single-arm, MDALS, and OALS designs. The round-trip arm lengths were chosen to be $33\,\mbox{s}$ and $32.4\,\mbox{s}$ for all three sensors for direct comparison purposes. There is no specific significance to the choice. The OALS sensor was computed with $\Delta t = 0.1\,\mbox{s}$ and perfect arm-length knowledge was assumed for both MDALS and OALS. When compared with the single-arm sensor, both MDLAS and OALS exhibit much flatter responses in the LISA measurement band. This allows arm-locking systems based on them to achieve more uniform suppression, particularly near the round-trip frequencies. The OALS has less ripple than the MDALS at frequencies below $\sim 100\,\mbox{mHz}$ but reaches a peak in-band ripple around $300\,\mbox{mHz}$ which is similar to that of MDALS. Both sensors can be used to build arm locking systems that meet the LISA performance criteria. The OALS is optimal in the sense that it is generated using optimal control theory techniques.  When paired with a suitable controller, we find that the net system performance is similar to that with achieved with the MDALS sensor, which gives us some confidence in that design.

\subsection{Controller Design}
 The second component in an arm locking system is a controller, which
takes the estimate of the laser frequency provided by the arm locking
sensor and generates a frequency tuning command for the laser. The
design goals of the controller are to provide sufficient gain within
the LISA measurement band to suppress the intrinsic frequency fluctuations
of the master laser (\ref{eq:MZnoise}) below the levels tolerated
by TDI (\ref{eq:preTDI-req}). As mentioned in section \ref{sec:Doppler},
care must also be taken to minimize the controller gain at very low
(below measurement band) frequencies to mitigate laser frequency pulling.

\begin{figure}[H]
\begin{centering}
\includegraphics[width=12cm]{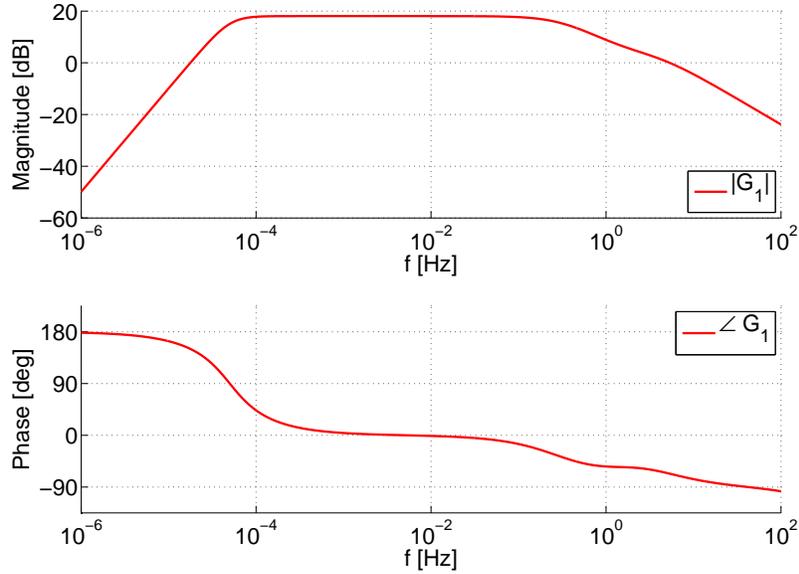}
\par\end{centering}
\caption{\label{fig:ControllerBode}Bode plot of arm locking controller}\end{figure}

The controller is based on a classical lead-lag design. It includes a second-order lead filter at 
the lower frequencies (break frequency at $0.05\,\mbox{mHz}$) to abate laser pulling due to 
uncompensated Doppler and Doppler derivative. It also includes a shaping filter and a single-order 
attenuation filter at $4\,\mbox{Hz}$ to limit the controller action to the LISA band. Figure \ref{fig:ControllerBode} 
contains a Bode plot of the controller.

\subsection{Simulation Design}

We implemented a discrete-time simulation of arm locking as described
in the preceding sections using the SIMULINK software package. Each
arm was modeled in a manner consistent with section \ref{sec:Background}.
The round-trip arm lengths were assumed to be $\tau_{12}+\tau_{21}=33\,\mbox{s}/c$
and $\tau_{13}+\tau_{31}=32.4\,\mbox{s}/c$, where $c$ is the speed
of light. The phase lock loops on $SC_{2}$ and $SC_{3}$ were assumed
to be perfect ($G_{2}=G_{3}\gg1$) with constant frequency offsets
($\nu_{M21}=10\,\mbox{MHz}$, $\nu_{M31}=15\,\mbox{MHz}$). The Doppler shifts in each arm
were modeled as a linearly-varying frequency with the coefficients
provided by the orbital model in Figure \ref{fig:Doppler} at a time $t=1\,\mbox{yr}$.
Doppler errors were linear in time with the coefficients provided in Table \ref{tab:DopErr}. The spectrum of intrinsic frequency fluctuations in the laser systems was modified from (\ref{eq:MZnoise}) to include two poles at $0.6\,\mu\mbox{Hz}$, limiting the total frequency excursion to $\sim(20\,\mbox{MHz})$ over the maximum simulation period of two weeks. A two pole roll off at $0.5\,\mbox{Hz}$ was added to the spacecraft jitter noise in (\ref{eq:JitterNoise}) to model the dynamics of the SC above the measurement band.

The simulation cadence was $500\,\mu\mbox{s}$. System dynamics, noise generators, and the controller
operated at this cadence. The OALS was implemented with the designed 10Hz sampling rate, with 
appropriate downsampling and upsampling filters providing the rate transitions. The OALS filter 
order was also reduced from the nominal order of 332 to 38 using balanced reduction\cite{Maciejowski89}. This 
reduction provides a dramatic increase in simulation speed without changing the behavior in the 
LISA measurement band.

\section{Results\label{sec:Results}}
\subsection{Component Noise Sources}

\begin{figure}[H]
\begin{centering}
\subfloat[\label{fig:AL_noise_freq}Frequency domain model]{\begin{centering}
\includegraphics[width=10cm]{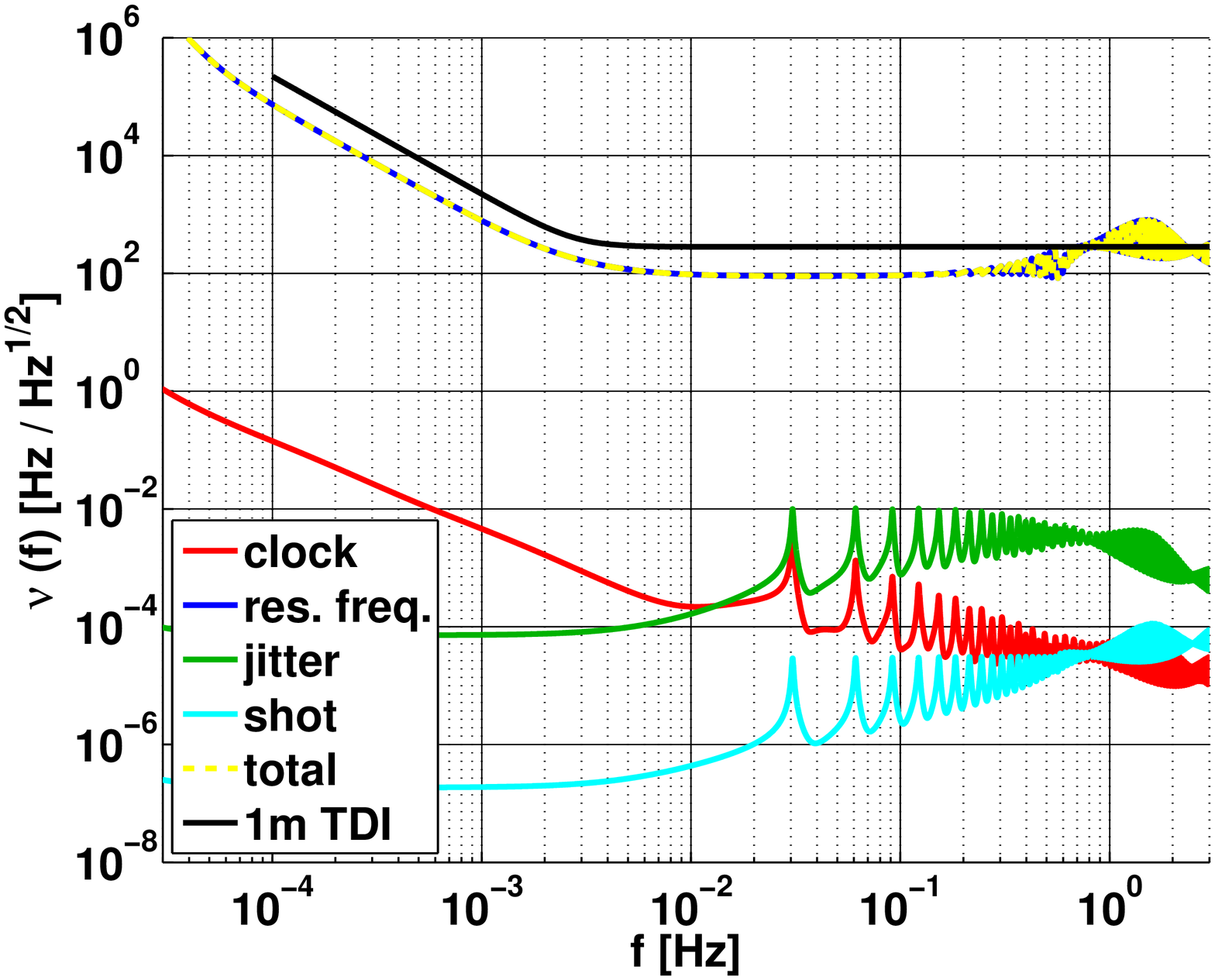}
\par\end{centering}}
\subfloat[\label{fig:AL_noise_time}Time domain model]
{\begin{centering}\includegraphics[width=10cm]{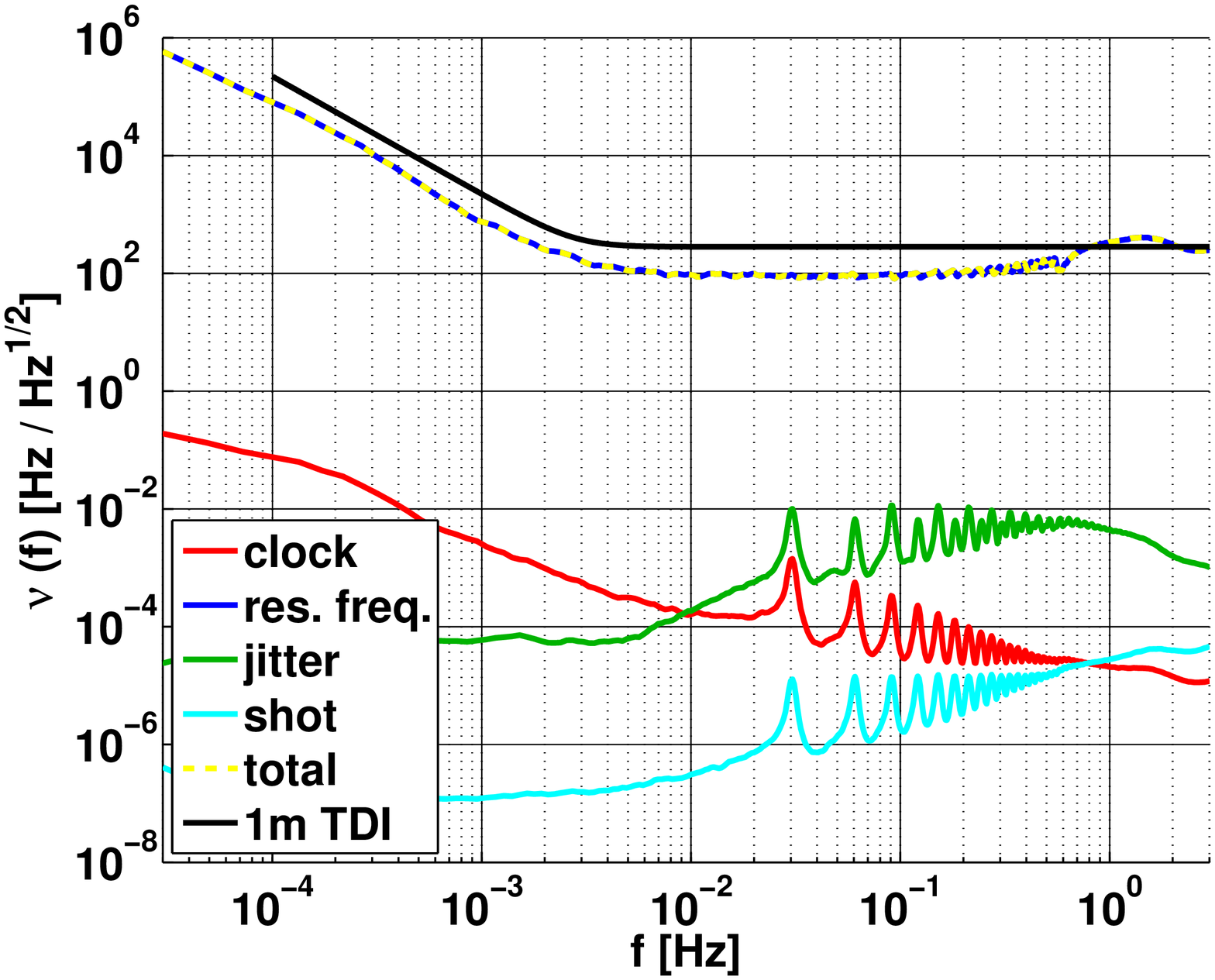}
\par\end{centering}}
\caption{\label{fig:AL_Noise_Project}Noise breakdown for arm locking system}
\end{centering}
\end{figure}

The first goal of the time-domain simulation was to verify the analytic,
frequency-domain model of the arm locking system. Figure \ref{fig:AL_noise_freq}
contains a noise decomposition of the OALS arm locking system derived
from an analytic model. As can be seen, the overall noise in the stabilized
laser is dominated by residual laser frequency noise, with the other noise sources
being nearly four orders of magnitude smaller. Figure \ref{fig:AL_noise_time}
shows a similar plot obtained using the time-domain simulations. To
obtain each curve, the simulation was run with all noise sources except
the source of interest turned off. In all cases, the Doppler estimation
errors were set to zero. The time series were then used to estimate
a spectra. The two plots show good agreement over most of the LISA
band. The primary differences are a broadening of the sharp spectral
features near $f=n/\tau$ and a roll-off at low frequencies in the
time-domain plots. Both of these effects are consistent with spectral
estimation errors in the logarithmic power spectral density algorithm \cite{Trobs_06} used to compute the spectra from the time series outputs.

\subsection{Laser Frequency Pulling}

\begin{figure}[H]
\begin{centering}
\includegraphics[width=12cm]{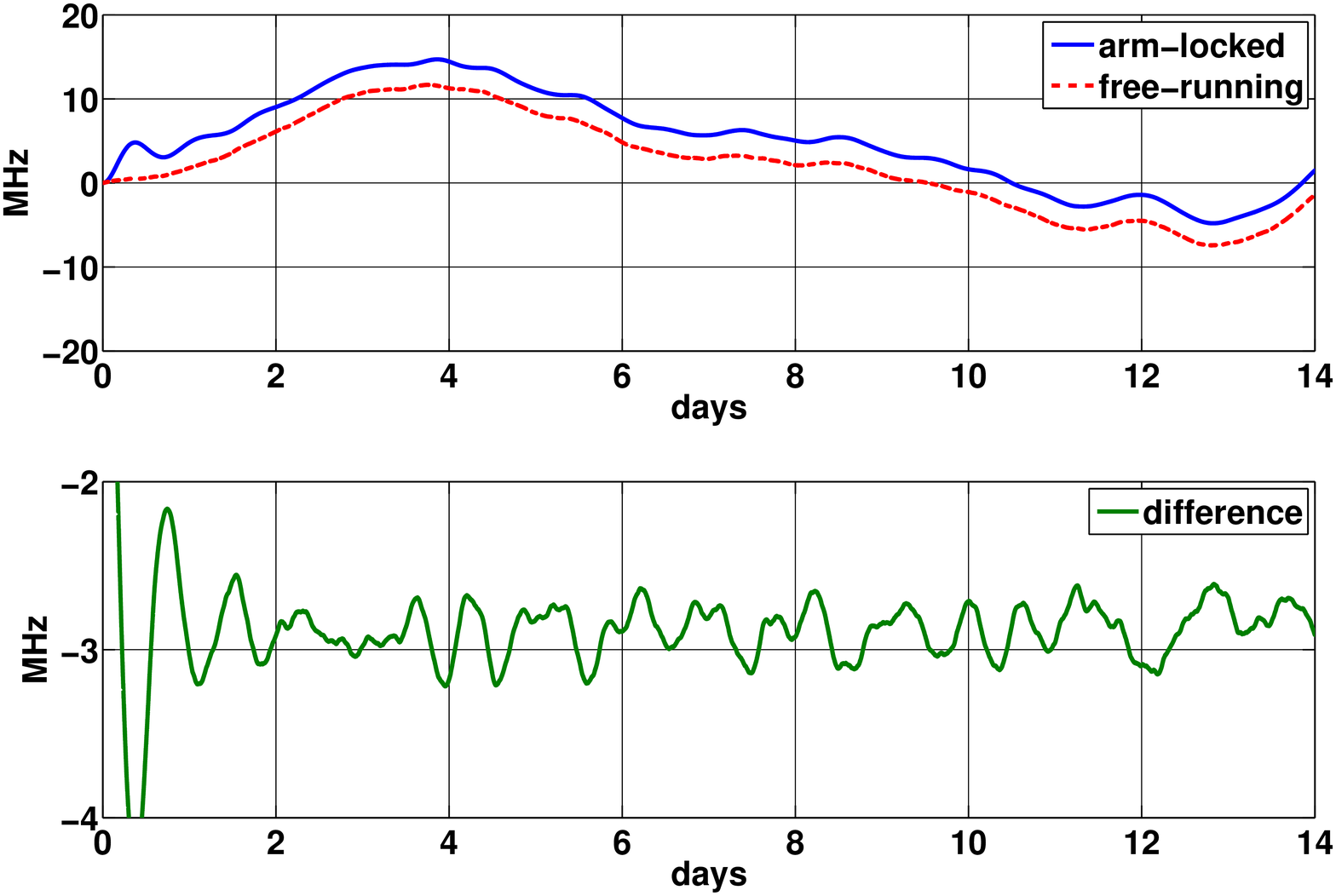}
\par\end{centering}

\caption{\label{fig:long_run_time}Top panel: comparison of arm locked laser
frequency drift with intrinsic laser frequency drift. Bottom panel:
Difference of the two curves in the upper panel giving a rough estimate of
the contribution to laser frequency drift from the arm locking system. }

\end{figure}

The second goal of the time-domain simulations was to explore phenomena
that are not easily treated analytically in the frequency domain.
The laser frequency pulling described in section \ref{sec:Doppler}
is an important example of such a phenomenon. We
ran a simulation spanning two full weeks (the expected time between
SC maintenance periods) with Doppler estimation errors consistent
with our models of the errors in the averaging method. Figure \ref{fig:long_run_time}
shows the results of this simulation. In the top panel, there are
two curves plotted: the frequency change of the arm locked system
and the frequency change of the intrinsic MZ noise. The first thing
to notice is that the arm locked system drifts over approximately
$20\,\mbox{MHz}$ over the two week simulation period, well within
the expected linear tuning range of the LISA lasers. The second thing to notice
is that the frequency drift in the arm locked system is approximately
equal to the drift in the intrinsic noise. This is due to the fact that the arm locking loop has no effect below the LISA measurement band. The lower panel
plots the difference of the arm locked and intrinsic frequency drifts, which
gives an indication as to the level of additional drift generated
by the arm locking system. After an initial transient decays over
the first few days, the remaining fluctuations are less than $1\,\mbox{MHz}$.
This demonstrates that this arm locking design does not produce any
significant pulling of the master laser frequency beyond what is already
present in the MZ stabilization system. 

\subsection{Robustness to Arm Length Errors}

Like the MDALS sensor, the OALS sensor requires some a priori knowledge
of the LISA arm lengths. Many of the same techniques described in section \ref{subsec:DoppMeth} that can potentially
be used for estimating the Doppler frequencies can also be applied to estimate
arm lengths. Which technique is most applicable will depend on how sensitive 
the performance of the arm locking system is to errors in the estimated arm
lengths used to compute the sensor. For example, if the maximum tolerable error is $\sim1\,\mbox{m}$ then active ranging is likely the best candidate. If, on the other hand, errors of $\sim10\,\mbox{km}$ are tolerable, it may be possible to compute them from orbital ephemerides on the ground and upload new coefficients to the OALS periodically. 

To check the robustness of the OALS to errors in the arm length, we first define the mean and differential arm lengths assuming $SC_1$ is the master SC,
\begin{equation}
\tau_{m}\equiv\frac{1}{2}\left[\tau_{12}+\tau_{21}+\tau_{13}+\tau_{31}\right],\label{eq:tau_m_def}
\end{equation}
\begin{equation}
\delta\tau\equiv\left[\tau_{12}+\tau_{21}-\tau_{13}-\tau_{31}\right],\label{eq:delta_tau_def}
\end{equation}
We then design an OALS for a specific set of nominal arm lengths, $\tau_m^{(0)}=32.85\,\mbox{s}$ and $\delta\tau^{(0)}=0.3\,\mbox{s}$, corresponding to the constellation geometry at $t\approx1.25\,\mbox{yrs}$ in the orbital solution used in Figure \ref{fig:Doppler}. This sensor is used to stabilize an array of arm locking systems with different true arm lengths, which corresponds not only to an error in the determination of the true arm lengths, but also represents the situation where the spacecraft constellation has evolved in time away from the design point. To quantify the effect of arm length errors, we define the figure of merit

\begin{equation} 
\Psi_{0}(\tau_m,\delta\tau)\equiv 20\,\log_{10}\left[\underset{f}{\max}\frac{\tilde{\nu}_{pre-TDI}(f)}{\tilde{\nu}_{O1}(f,\tau_m,\delta\tau)}\right]\label{eq:costFcn0}\end{equation}

where $\tilde{\nu}_{O1}(f,\tau_m,\delta\tau)$ is the residual noise in the master laser and $\tilde{\nu}_{pre-TDI}(f)$ is the residual noise requirement specified in (\ref{eq:preTDI-req}). $\Psi_0$ measures the minimum margin in the LISA measurement band between the arm locking system system and the pre-TDI requirement. Where this minimum is positive, the frequency stabilization is guaranteed to meet performance requirements.

\begin{figure}[H]
\begin{centering}
\includegraphics[width=12cm]{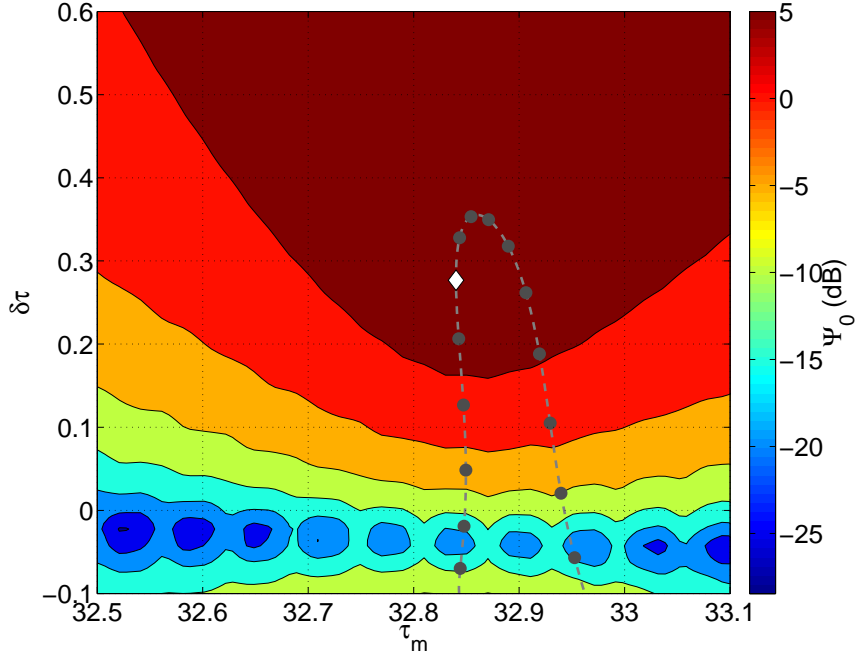}
\caption{\label{fig:armSens}Robustness of an example OALS-based arm locking system to errors in arm length versus the mean ($\tau_m$) and differential ($\delta\tau$) arm lengths. The contours show $\Psi_0$, the minimum margin in any given frequency bin within the LISA measurement band. The white diamond marks the design point for the sensor and the dashed line shows the evolution of the arm lengths due to LISA orbital motion near the design point with the grey dots spaced in time by 15 days.  The dark red color shows a positive margin between 0 and 5 dB indicating that the performance of the system meets or exceeds the requirement.  Evolution of the system is from left to right, indicating that the system has adequate performance from approximately 20 days before the design point to approximately 100 days after.}
\end{centering}
\end{figure}

Figure \ref{fig:armSens} shows a contour plot of $\Psi_0$ for this example plotted on the $(\tau_m,\delta\tau)$ plane. The design point $(\tau_m^{(0)},\delta\tau^{(0)})$ is indicated by a white diamond. The evolution of $\tau_m$ and $\delta\tau$ due to the LISA orbit near the design is indicated by the dashed line with the grey dots indicating time intervals of 15 days. Figure \ref{fig:armSens} shows that positive margin exists for $\sim 20\,\mbox{days}$ prior to the design point and $\sim 100\,\mbox{days}$ afterwards. This is much larger than the expected intervals between maintenance activities, indicating that updating the OALS sensor coefficients will not drive the maintenance schedule of the mission. The existence of a large operating window gives us some confidence that the OALS design is robust enough for the LISA application.

When arm length errors eventually do cause the system performance to degrade to the point where the margin is inadequate, it will be necessary to change the OALS coefficients and possibly which SC is designated as the master.  Changing the coefficients can likely be done smoothly without losing lock or degrading system performance. Changing the master SC will require re configuring of the phase lock loops aboard all SC and will result in some down time. This should only be required when $\delta\tau$ for a certain arm combination becomes sufficiently small, likely $1-3$ times per year.

\section{\label{sec:Conclusions}Conclusions}
Arm Locking is a candidate laser frequency stabilization technique for LISA. We have used a time domain simulation to study a design for an arm locking system based on a Kalman filter optimal blended sensor and a controller which meets the frequency stability requirements for LISA assuming the master laser is pre-stabilized to a level of $800\,\mbox{Hz}/\sqrt{\mbox{Hz}}$.  Time domain simulations allowed us to study transient phenomena and the performance of the stabilization system as the conditions of the LISA constellation evolve during the normal orbital motion, including Doppler shifts and imperfectly known arm lengths.  The simulations indicate that it is possible to implement arm locking without excessive pulling of the the master laser frequency, and that the arm locking sensor performance is robust against errors in the absolute arm length estimates. This robustness allows the sensor to be periodically updated with pre-computed filter coefficients at intervals that are operationally reasonable.
Although our time-domain simulation necessarily included specific sensor and controller designs, the same simulation infrastructure could be applied to study other candidate sensor and controller designs.

\begin{acknowledgments}
We would like to thank Kirk McKenzie for providing
the tools needed to compute the MDALS arm-locking performance and Steve Hughes
for providing LISA orbital data.
\par
Copyright (c) 2011 United States Government as represented by the Administrator of the National Aeronautics and Space Administration. No copyright is claimed in the United States under Title 17, U.S. Code. All other rights reserved.
\end{acknowledgments}

\appendix
\section{Key to Notation\label{sec:notationKey}}
\begin{table}[H]
\caption{\label{tab:NotationTable}Partial key to notation and comparison with \cite{McKenzie_09}}
\begin{tabular}{|c|c|c|}
\hline 
Symbol & Description & Correspondence in \cite{McKenzie_09}\tabularnewline
\hline
\hline 
$G_{i}(s)$ & Controller transfer function on $SC_{i}$ & $G_{i}(s)$\tabularnewline
\hline 
$\textbf{S}$ & Arm locking sensor vector & $\textbf{S}$\tabularnewline
\hline 
$\overrightarrow{V}_{i}$ & total velocity of $SC_{i}$ & N/A\tabularnewline
\hline 
$\overrightarrow{V}_{Oi}$ & orbital component of $SC_{i}$ velocity & N/A\tabularnewline
\hline 
$\delta\overrightarrow{V}_{i}$ & jitter component of $SC_{i}$ velocity & $\left(\delta\overrightarrow{V}_{i}-\delta\overrightarrow{V}_{j}\right)\cdot\hat{\eta}_{ij}=\frac{\partial}{\partial t}\Delta X_{ij}$\tabularnewline
\hline 
$\tilde{y}_{i}(f)$ & fractional frequency fluctuations of $SC_{i}$ clock & $\tilde{y}_{i}(f)$\tabularnewline
\hline 
$\hat{\eta}_{ij}$ & unit vector from $SC_{i}$ to $SC_{j}$ & N/A\tabularnewline
\hline 
$\lambda$ & wavelength of lasers & $\lambda$\tabularnewline
\hline 
$\nu_{Aij}$ & output of frequency meter $ij$ & $\frac{\partial}{\partial t}\phi_{Aij}$\tabularnewline
\hline 
$\nu_{B1}$ & output of arm locking sensor  & $\frac{\partial}{\partial t}\phi_{B1}$\tabularnewline
\hline 
$\nu_{Cij}$ & clock noise generated by frequency meter $ij$ & $\frac{\partial}{\partial t}\phi_{Cij}$\tabularnewline
\hline 
$\nu_{Dij}$ & Orbital Doppler shift measured by frequency meter $ij$ & N/A\tabularnewline
\hline 
$\nu_{E1j}$ & Error in heterodyne model  & $\nu_{DE1j}$\tabularnewline
\hline 
$\nu_{Hij}$ & heterodyne signal at photoreceiver $ij$ & N/A\tabularnewline
\hline 
$\nu_{Jij}$ & Spacecraft jitter Doppler shift at measurement $ij$ & $\frac{\partial}{\partial t}\phi_{Xij}$\tabularnewline
\hline 
$\nu_{Li}$ & intrinsic frequency noise of Laser $i$ & $\frac{\partial}{\partial t}\phi_{Li}$\tabularnewline
\hline 
$\nu_{Mij}$ & Heterodyne model signal on frequency meter $ij$ & $\Delta_{i1}\:\:i=2,3$\tabularnewline
\hline 
$\nu_{M+(-)}$ & model of common (differential) component of heterodyne signals on
$SC_{1}$ & N/A\tabularnewline
\hline 
$\nu_{0+(-)}$ & constant part of $\nu_{M+(-)}$ & $\nu_{0+(-)}$\tabularnewline
\hline 
$\gamma_{0+(-)}$ & linear part of $\nu_{M+(-)}$ & $\gamma_{0+(-)}$\tabularnewline
\hline 
$\alpha_{0+(-)}$ & quadratic part of $\nu_{M+(-)}$ & $\alpha_{0+(-)}$\tabularnewline
\hline 
$\nu_{Oi}$ & frequency output of Laser $i$ & $\frac{\partial}{\partial t}\phi_{Oi}$\tabularnewline
\hline 
$\nu_{Sij}$ & shot noise at photoreceiver $ij$ & $\frac{\partial}{\partial t}\phi_{Sij}$\tabularnewline
\hline 
$\tau_{ij}$ & light travel time from $SC_{i}$ to $SC_{j}$ & $\tau_{ij}$\tabularnewline
\hline
\end{tabular}
\end{table}

\bibliographystyle{prsty}
\bibliography{Thorpe041811}

\begin{thebibliography}{10}

\bibitem{Bender_98}
P. Bender, K. Danzmann, and the LISA Study~Team, Technical Report No.~MPQ233,
  Max-Planck-Institut fur Quantenoptik, Garching (unpublished).

\bibitem{Jennrich_09}
O. Jennrich, Class. Quant. Grav. {\bf 26},    (2009).

\bibitem{Armstrong_99}
J. Armstrong, F. Estabrook, and M. Tinto, The Astrophysical Journal {\bf 527},
  814  (1999).

\bibitem{Shaddock_03}
D. Shaddock, M. Tinto, F. Estabrook, and J. Armstrong, Physical Review D {\bf
  68},    (2003).

\bibitem{Shaddock_09}
D. Shaddock and et~al., LISA Project technical note LISA-JPL-TN-823, Laser
  Interferometer Space Antenna (unpublished).

\bibitem{Thorpe10}
J. Thorpe, Class. Quant. Grav. {\bf 27},    (2010).

\bibitem{Sheard_03}
B. Sheard, M. Gray, D. McClelland, and D. Shaddock, Physics Letters A {\bf
  320},  9  (2003).

\bibitem{Sutton_08}
A. Sutton and D. Shaddock, Physical Review D {\bf 78},    (2008).

\bibitem{McKenzie_09}
K. McKenzie, R.~E. Spero, and D.~A. Shaddock, Physical Review D {\bf 80},
  (2009).

\bibitem{Sheard_05}
B. Sheard, M. Gray, D. Shaddock, and D. McClelland, Class. Quant. Grav. {\bf
  22},    (2005).

\bibitem{Marin_05}
A.~G. Marin {\it et~al.}, Class. Quant. Grav. {\bf 22},  S235  (2005).

\bibitem{Wand_09}
V. Wand {\it et~al.}, Journal of Physics: Conference Series {\bf 154},
  (2009).

\bibitem{Shaddock_08}
D. Shaddock, Class. Quant. Grav. {\bf 25},    (2008).

\bibitem{Shaddock_06}
D. Shaddock {\it et~al.},  in {\em Proceedings of the Sixth International LISA
  Symposium}, edited by S. Merkowitz and J. Livas (AIP Conference Proceedings,
  ADDRESS, 2006), Vol.~873, pp.\ 654--660.

\bibitem{Steier_09}
F. Steier {\it et~al.}, Class. Quant. Grav. {\bf 26},    (2009).

\bibitem{Armano_09}
M. Armano and et~al., Class. Quant. Grav. {\bf 26},    (2009).

\bibitem{Klipstein_06}
W. Klipstein {\it et~al.},  in {\em Laser Interferometer Space Antenna: 6th
  International LISA Symposium}, edited by S. Merkowitz and J. Livas
  (PUBLISHER, ADDRESS, 2006), Vol.~873, pp.\ 312--318.

\bibitem{Maghami03}
P. Maghami and T. Hyde, Class. Quant. Grav. {\bf 20},  S273  (2003).

\bibitem{Hughes_08}
S. Hughes, Journal of Guidance, Control, and Dynamics {\bf 31},    (2008).

\bibitem{Thornton05}
C. Thornton and J. Border, {\em {Radiometric Tracking Techniques for Deep-Space
  Navigation}}, {\em Deep-space communications and navigation series} (John
  Wiley \& Sons, ADDRESS, 2005).

\bibitem{Esteban_09}
J.~E. Delgado {\it et~al.}, Journal of Physics: Conference Series {\bf 154},
  (2009).

\bibitem{Heinzel11}
G. Heinzel, Class. Quant. Grav. {\bf 28},    (2011).

\bibitem{Maghami_09}
P. Maghami, J. Thorpe, and J. Livas, {\em Proceedings of the SPIE Symposium on
  Advanced Wavefront Control:methods, Devices, and Applications VIII}
  (PUBLISHER, ADDRESS, 2009), Vol.~7466, pp.\ M1--M11.

\bibitem{Stengel}
R. Stengel, {\em Stochastic Optimal Control: Theory and Application} (John
  Wiley \& Sons, New York, NY, 1986).

\bibitem{Phillips_and_Nagle}
C. Phillips and H. Nagle, {\em Digital System Analysis and Design}, 3rd ed.
  (Prentice Hall, Englewood Cliffs, NJ, 1995).

\bibitem{Maciejowski89}
J. Maciejowski, {\em Multivariable Feedback Design} (Addison-Wesley, ADDRESS,
  1989).

\bibitem{Trobs_06}
M. Tr$\ddot{o}$bs and G. Heinzel, Measurement {\bf 39},  120  (2006).

\end{thebibliography}

\end{document}